\documentclass{kluwer} 
\usepackage{psfig}
\begin{document}                                                                                   
\begin{article}
\begin{opening}         
\title{Molecular Gas in Galaxies}

\author{Francoise \surname{Combes}}  
\runningauthor{F. Combes}
\runningtitle{Molecular Gas in Galaxies}
\institute{DEMIRM, Observatoire de Paris, \\
61 Av. de l'Observatoire, F-75 014, Paris, France}
\date{July 14, 2000}

\begin{abstract}
Knowledge of the molecular component of the ISM is fundamental to 
understand star formation. The H$_2$ component appears to
dominate the gas mass in the inner parts of galaxies,
while the HI component dominates in the outer parts.
Observation of the CO and other lines in normal and starburst 
galaxies have questioned the CO-to-H$_2$ conversion factor, and 
detection of CO in dwarfs have shown how sensitive the conversion f
actor is to metallicity. Our knowledge has made great 
progress in recent years, because of sensitivity and spatial resolution 
improvements.  Large-scale CO maps of nearby galaxies are now available, 
which extend our knowledge on global properties, radial gradients, and spiral 
structure of the molecular ISM. Millimetric interferometers reveal high 
velocity gradients in galaxy nuclei, and formation of embedded structures, 
like bars within bars.  Galaxy interactions are very effective to enhance
gas concentrations and trigger starbursts. Nuclear disks or rings
are frequently observed, that concentrate the star formation activity.
Since the density of starbursting galaxies is strongly increasing with
redshift, the CO lines and the mm dust emission are a privileged tool to 
follow evolution of galaxies and observe the ISM dynamics at high redshift: 
they could give an answer about the debated question of the star-formation 
history, since many massive remote starbursts could be dust-enshrouded.
\end{abstract}
\end{opening}

\vspace*{-0.5cm}
\section{Introduction}

The interstellar medium, according to its density and physical
conditions, can be found essentially as atomic hydrogen or molecular
hydrogen. The latter plays a fundamental role in star formation.
But the bulk of molecular hydrogen is cold ($\sim$ 10K), does not 
radiate and is thus completely invisible. The H$_2$ component is known 
in galaxies essentially from the CO tracer, but the way to derive the total
amount of molecules is uncertain, and mapping other tracers
is of prime importance: UV absorption lines, dust emission, 
mid-IR rotational lines of warm H$_2$, etc...

This paper reviews all of our indirect knowledge about the H$_2$
component, and compares all the tracers in order to determine
how much molecular mass is in galaxies. The relation with
the atomic gas HI is described. The molecular content of galaxies
is traced as a function of morphological type, of evolution state.
Its role is emphasized in dynamics of galaxies (bars and spirals)
and in galaxy interactions and mergers. Finally, the H$_2$ content
as a function of redshift is briefly discussed, 
as a way to trace the evolution of star formation, and
to determine the importance of starbursts versus
AGN for instance.

\section{CO to H$_2$ conversion ratio}  

\subsection{UV absorption lines}

The CO molecule is excited by H$_2$ collisions, and should be a
good tracer of molecular gas; but its main rotational lines are most 
of the times optically thick. It is possible to observe its isotopic
substitutes $^{13}$CO or C$^{18}$O, but these are poor tracers
since they are selectively photo-dissociated, and trace only the
dense cores.

The H$_2$/CO conversion ratio was first calibrated by 
comparing the UV absorption lines of CO and H$_2$ along the same line
of sight (Copernicus, e.g. Spitzer \& Jenkins 1975;
 ORFEUS, cf Richter et al., 1999a,b). This is now becoming possible
at much larger-scale, with the FUSE satellite, and molecular hydrogen 
bands have been observed toward several stars lying behind diffuse and
translucent clouds (Tumlinson et al. 1999, Snow 2000).
However, only very low column densities are accessible, 
in order to see the background source, and therefore these observations 
sample only the diffuse gas, which is not representative of the 
global molecular component. It is well known now that the 
conversion factor might vary by one order of magnitude
from diffuse to dense clouds, since the relation between the virial
mass and CO luminosity is non linear.

\subsection{Virial hypothesis}

The main justification to use an H$_2$/CO conversion ratio is
the Virial hypothesis: in fact, the CO profiles do not yield the column
densities, but they give the velocity width $\Delta$V of molecular clouds.
Once the latter are mapped, and their size R known, the virial mass can
be derived, proportional to $\Delta$V$^2$ R. There exists a good relation
between the CO luminosity and the virial mass; however it is a power-law 
of slope different from 1: M$_{\rm V} \propto$ L$_{\rm CO}^{0.76}$
(cf Solomon et al. 1987). Therefore the conversion ratio should vary by 
more than a factor 10 from small to Giant Molecular Clouds (GMC).
At large-scale in our Galaxy, and in external galaxies, the observations 
provide an average over the whole mass spectrum of clouds, and 
the hypothesis is made that this average conversion ratio is the same
from galaxy to galaxy. If T$_b$ is the brightness temperature
of the average cloud, the conversion ratio X should vary as n$^{1/2}$/T$_b$,
 where n is the average density of the cloud. This does not take into
account the influence of the gas metallicity.

\subsection{Variation with metallicity: dwarfs and LSBs}

At constant H$_2$ column density, the CO luminosity varies
with the metallicity Z, sometimes more than linearly. In the
Magellanic Clouds, LMC or SMC (Rubio et al 1993), the conversion
ratio X might be 10 times higher than the "standard" ratio. The ratio
can be known for local group galaxies, since individual clouds can
be resolved, and virial masses computed (Wilson 1995).

The strong dependency of the H$_2$/CO conversion ratio on
metallicity Z is also the main problem in the observations
of dwarf and Low Surface Brightness (LSB) galaxies. Both have 
low metallicity. Not only, the low abundance of C and O lowers 
the abundance of CO, but also the dust is less abundant,
and therefore the UV light is less absorbed, and
spread all over the galaxy, photo-dissociating the CO molecules.
When the dust is depleted by a factor 20, there should be only
 10\% less H$_2$, but 95\% less CO (Maloney \& Black 1988).

CO emission is in general very low in dwarf galaxies, 
and it is difficult to know their H$_2$ content.
If the HI/H$_2$ ratio is assumed constant from
galaxy to galaxy, then
X varies with Z$^{-2.2}$ (Arnault et al 1988).
Recent results by Gondhalekar et al (1998), 
Taylor et al (1998) and Barone et al (2000),
confirm this strong dependency on metallicity,
increasing sharply below 1/10th of solar metallicity.

Low-surface-brightness galaxies have large characteristic radii, 
large gas fraction and are in general dark matter dominated;
they are quite un-evolved objects. Their total gas content is 
similar to that of normal galaxies (McGaugh \& de Blok 1997).
But CO is not detected in LSB (de Blok \& van der Hulst 1998,
Braine et al. 2000a).Due to their low surface density, below 
the threshold for star formation, these galaxies have a very
low efficiency of star formation (Van Zee et al 1997).
The cause could be the absence of companions, since
LSB live in poor environments (Bothun et al.  1993).
It is well known that galaxy interactions, by driving in
a high amount of gas, trigger star formation.

\vspace*{-0.5cm}
\section{Other promising tracers}

\vspace*{-0.5cm}
\subsection{Dust as a tracer}

\begin{figure}
\centering
\psfig{figure=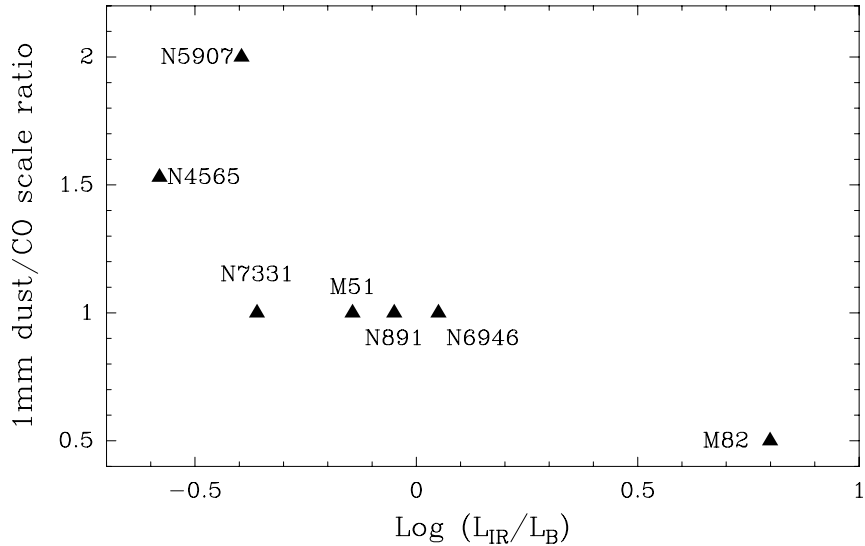,bbllx=3cm,bblly=0cm,bburx=20cm,bbury=27cm,width=12cm,angle=-90}
\caption{Scale-length ratio between the mm dust emission and the CO(1-0) emission
versus the L$_{IR}$/L$_B$ ratio, for NGC 891 (Gu\'elin et al. 1993), M82 (Thuma et al. 2000),
NGC 4565 (Neininger et al. 1996), M51 (Gu\'elin et al. 1995), NGC 5907 (Dumke et al. 1997),
NGC 6946 (Bianchi et al. 2000) and NGC 7331 (Bianchi et al. 1998). The scales have been 
compared, when the intensity has been divided by e with respect to the center.}
\label{dust}
\end{figure}

At millimetric wavelengths, in the Rayleigh-Jeans domain,
dust emission depends linearly on temperature, 
and its great advantage is its optical thinness.
In some galaxies, CO and dust emission fall similarly with radius, like
in NGC 891 (Gu\'elin et al. 1993). In other, such as NGC 4565 (Neininger
et al. 1996), the dust emission falls more slowly than CO, although more
rapidly than HI emission. This can be interpreted by the exponential decrease
of metallicity with radius. The dust/HI ratio follows this dependency, while
CO/HI is decreasing more rapidly (either due to metallicity, or excitation
problems). In M82, due to the intense starburst and related cosmic
ray heating, the CO is much more extended than the dust 1.2mm emission
(Thuma et al. 2000). Figure \ref{dust} compares the CO and mm dust
emission scale lengths in a few spiral galaxies, as a function of star-formation 
activity (traced by far-infrared to blue luminosity ratio): in galaxies with
active star formation, the CO emission is enhanced (the derived M(H$_2$)/M(HI)
is larger), and the dust closely follows the CO radial distribution.
On the contrary, in less active galaxies, the HI mass dominates, and 
the dust emission follows the HI in the outer parts of the galaxy disk,
beyond the end of the CO disk.
ISO 200$\mu$m images have shown that the dust radial distribution 
is more extended than stellar disks (Alton et al. 1998), and that the
ratio of cold to warm dust is increasing with radius. 

\subsection{Rotational mid-IR lines}

A very small fraction of the molecular gas can
be excited to very high temperatures through shocks and
then be observed directly, through the ro-vibration lines.
Starbursts and mergers reveal strong
2.2 $\mu$m emission, like in NGC 6240 (DePoy et al 1986).
The source of excitation has long been debated (X-ray heating,
UV fluorescence, shocks...) and it was recently concluded that
global shocks were responsible (van der Werf et al. 1993, Sugai et al. 1997).
Pure rotational lines have been observed with ISO. In
Arp220, as much as 10\% of the ISM could be in the
warm phase, i.e. 3 10$^9$ M$_\odot$ (Sturm et al. 1996)
while CO observations conclude to a total M(H$_2$) = 3.5 10$^{10}$
M$_\odot$ (Scoville et al. 1991). In normal galaxies, the warm H$_2$ could be
 less abundant (Valentijn et al. 1996). 

The ISO satellite also allowed to explore the pure H$_2$ rotational lines,
the first S(0) (J= 2--$>$ 0) of the para-hydrogen having its upper level
at $\sim$ 500 K above ground (28 $\mu$m in wavelength).
Valentijn \& van der Werf (1999) derived the radial distribution
of the two first lines in the edge-on galaxy NGC 891. Surprisingly,
the S(0) emission has a rather flat distribution, while the CO
emission falls down exponentially. The different line-widths
for the two first lines (the S(1) being much narrower) tend to support
an interpretation in terms of a two-component medium, where
a cool H$_2$ gas dominates the S(0) emission. The temperature is
then below 90K, and the derived column density of H$_2$ is
10$^{23}$ cm$^{-2}$, ten times larger than the HI column density.
This is based on the assumption that the ortho-para ratio is about
1. The mass derived is then sufficient to explain the flat rotation 
curve. Other choices of the parameters could reduce the derived
gas mass, however. How is this H$_2$ gas heated? The ionising
and photo-dissociating radiation from stars are not sufficient in the
outer parts of the galaxy disk. It is possible that the usual 
turbulence of molecular clouds, maintained by gravitational instabilities,
is producing mild shocks, sufficient to heat a fraction of the 
molecular gas, at relatively low temperatures. This fraction of
warm H$_2$ might then be a good tracer of the bulk of cold H$_2$
in the absence of CO molecules.

\section{ Comparison with the HI component}
\subsection{Radial distribution}

The differences between HI and H$_2$ (or CO) radial
distributions in galaxies is striking: while the
N(H$_2$)/N(HI) in the center can reach 10 or 20, it
falls below 1 and even 0.1 in the outer parts. 
While all components related to star formation,
the blue luminosity from stars, the H$\alpha$ (gas ionised
by young stars), the radio-continuum (synchrotron related
to supernovae), and even the CO distribution, follow
an exponential distribution, the HI gas alone is
extending much beyond the ``optical'' disk, sometimes
in average by a factor 2 to 4 (R$_{HI}$ = 2-4 R$_{opt}$).
The HI gas has very often a small deficiency in the center.
 Would this mean that the atomic gas is transformed in
molecular phase in the denser central parts? This is
possible in some galaxies, where the HI and CO distribution
appear complementary, but it is not the general case,
all possibilities have been observed, including a central
gaseous depletion, both in CO and HI (like in M31 or NGC 7331).

Smith et al. (2000) have recently proposed a new probe of
H$_2$ in galaxies. Considering that the HI gas is coming
from dissociated molecular gas in PDRs, the volumic density
of local H$_2$ can be deduced from measurements of the HI
column density together with the far-ultraviolet (FUV) photon flux.
They apply this idea to M101, and find that, after correction for the 
metallicity gradient and for the extinction of the FUV emission, the H$_2$
density is about constant over radius up to 26 kpc from the center,
i.e. close to  R$_{25}$.

\subsection{Vertical structure}
In our own Galaxy, and in external galaxies seen edge-on,
the galaxy disks appear much narrower in CO emission
than in HI. This suggests that the molecular gas is more
confined to the plane, due to a much lower 
vertical velocity dispersion. Surprisingly, this is not the 
case: in face-on galaxies both CO (Combes \& Becquaert 1997) 
and HI (Kamphuis 1992) velocity dispersions
are observed of similar values ($\sigma_v \sim 6$ km/s),
and remarkably constant with radius. 
A possible interpretation is that both gas are 
the same dynamical component, which changes
phase along its vertical oscillations. It is possible that
the H$_2$ gas follows the HI, but the CO is photo-dissociated
at high altitudes, or not excited. Or even the H$_2$ could
disappear, since the chemistry time-scale  ($\sim$ 10$^5$ yr)
is much smaller than the dynamical z-time-scale
($\sim$ 10$^8$ yr).

\section{CO and H$_2$ content as a function of type}

From the Amherst CO survey of more than 300 galaxies, 
Young \& Knezek (1989) and Young \& Scoville (1991) have
concluded that the average molecular content was
comparable to the atomic content:  M(H$_2$)/M(HI)  $\sim$ 1.
However, some of these galaxies were selected from their
IRAS flux, and this could introduce a bias.
A recent survey by Casoli et al (1998) near
the Coma cluster has shown an average
M(H$_2$)/M(HI)  $\sim$ 0.2.

It is well established that the HI component is proportionally
more abundant relative to the total mass in late-type
galaxies. The opposite trend is observed for the H$_2$,
at least as traced by the CO emission.
M(H$_2$)/M(HI) is therefore smaller for late-types,
by a factor $\sim$ 10. However, this
could be entirely a metallicity effect. Since the
metallicity is increasing with the mass of the galaxy,
a test is to select the most massive galaxies of
late-type. For these high-mass galaxies, there is no trend of decreasing
H$_2$ fraction with type (Casoli et al. 1998).

\section{Role in Dynamics}

\subsection{Bars, nuclear bars}

The dissipative character of the gas is fundamental for the formation
of bars within bars, and for the transfer of angular momentum
to the outer parts, to allow the radial inflows. The atomic gas is
most of the time depleted in galaxy centers, and the molecular
component is the best tool to trace the gas behaviour there.
The millimeter interferometers provide now a sufficient 
spatial resolution for the CO maps.  
In general, barred galaxies show characteristic features corresponding
to the offset dust lanes seen in optical. When these two features
are seen only to start from the nucleus, they are called twin-peaks
(Kenney et al. 1992) and correspond to the presence of an inner Lindblad 
resonance, implying orbits perpendicular to the bar in the center.
The gas is often concentrated in resonance rings (nuclear rings,
cf Sakamoto et al 1999, Thornley et al. 1999), or in nuclear spirals.
 It is often difficult to discriminate between several possibilities
to account for the non-circular motions observed: nuclear bars
or warps, as in the Seyfert galaxies NGC 1068 and NGC 3227
(Schinnerer et al. 2000).

\subsection{Gas in shells}

Molecular gas can sometimes be detected far from the galaxy
centers, outside the optical image of the galaxy. This is the case
for the shells in Centaurus A (Charmandaris, Combes \& van der Hulst 2000). 
Shells are formed by stars of disrupted companions, by a phase-wrapping 
process (Quinn 1984). A large fraction of elliptical galaxies possess
shells (Schweizer \& Seitzer 1992), and this is believed to
support the hierarchical merging scenario for their formation.
In a merging event, gas is expected to dissipate and fall to the 
center (Weil \& Hernquist 1993). However, atomic gas has been 
observed associated with shells (Schiminovich et al 1994).
 In the phase-wrapping process, this is only possible if there
exists a gas component condensed in small clouds, with
a large mean free path. This component has only very small 
dissipation, and behaves more like ballistic particles, like stars.
 The ensemble of molecular clouds has such properties.
If the disrupted companion possessed dense molecular clouds,
they could have followed the stars in the shell formation, 
and through photo-dissociation and evaporation, reform 
some atomic gas in shells. The detection of CO emission in shells support
this scenario  (Charmandaris et al. 2000). The surprise if the derived large
amount of molecular gas in shells: 50\% of gas in shells is molecular,
and more than 10\% of all the gas in Centaurus A is away from 
the inner parts. Moreover, the H$_2$/HI ratio is the same in the
nuclear disk and in the shells. How has the gas been enriched in 
metals, so far from the nucleus? The solution might lie in the 
recent star formation triggered in the shell gas by the impact of the 
radio jet (Graham 1998, 99). The shells detected in CO are precisely
aligned with the radio jet, and the recently formed stars could have
enriched the observed gas in metals and account for the CO detection.

\subsection{Tidal dwarfs}

Tidal dwarfs are small systems becoming gravitationally
bound within tidal tails dragged by the interaction between
two massive gas-rich galaxies. The collapse of the gas
in these systems trigger new star formation 
(e.g. Duc \& Mirabel 1998).

Braine et al (2000) report the discovery of CO emission
in two tidal dwarf galaxies, in the Arp105 and Arp245 systems. 
In both cases, they derive that the molecular gas 
peaks at the same location as the HI gas, and infer from this
 that the molecular gas formed from the atomic hydrogen,
rather than being torn in molecular form from the interacting galaxies.
In fact, this could also be a consequence of the CO being visible
in these dwarfs, only because of the metallicity enrichment due
to the new stars formed there (see previous section).

\section{ CO at high redshift}

The recent years have seen the rapid development of sub-mm surveys
in blank areas, searching for high-redshift continuum sources. Since the
 spectral energy distribution (SED) of starbursting galaxies have
a characteristic peak around 60-100 $\mu$m
due to dust heated by newly born stars, the millimeter domain
becomes a privileged range to detect these objects at $z$ up to 10.
The slope of the SED (in $\nu ^4$)  is such that the K-correction is even
negative, i.e. it is more easy to detect objects at higher redshift than
$z = 1$, at a given frequency, and sky surveys could be dominated
by remote objects (see e.g. Blain \& Longair \cite*{blain93, blain96}). 
The density of sources detected up to now account for a significant 
fraction of the CIBR (Hughes et al. 1998).
Identification of the sources (redshifts) and of the nature of the emission
is difficult. At least 20\% of the sources reveal an
AGN activity, and most of them  are at relatively
low redshift $ 1 < z < 3 $ (Barger et al. \cite*{barger99}).

The detection of large amounts of molecular gas could help
to identify starbursts versus AGNs. However, the detection 
of the CO lines are much more difficult, since the K-correction
is not negative (Combes et al 1999). Today, a dozen of sources have 
been detected in CO at redshifts between 2 and 5, and most of
them are amplified by gravitational lensing. With the 
new millimeter instruments planned over the world
(the Green-Bank-100m of NRAO, the LMT-50m of UMass-INAOE,
the ALMA (Europe/USA) and the LMSA (Japan) interferometers)
the sensitivity will be enhanced such as to detect most of the
sources identified in the continuum.
 This will bring fundamental information about the cold gas component
in high-z objects and therefore about the physical conditions of
the formation of galaxies and the first generations of stars.
At high enough redshifts, most of the galaxy mass could be molecular.
The starburst occuring in these objects could enrich quickly the 
ISM to solar values (Elbaz et al. 1992).

\vspace*{-0.5cm}
\section{Conclusion}

Our knowledge of the molecular component of galaxies is improving 
fast, and it is now realized how much the H$_2$/CO conversion ratio
is varying with type and star forming activity. Other tracers will
be highly valuable in the near future: mm dust emission and
pure H$_2$ rotational lines.  The CO tracer is complementary to the
HI line to trace the gas dynamics in galaxies, since their radial distribution
are quite different and anti-correlated. With improved sensitivity,
it is now possible to detect CO lines even outside the optical galaxies.
The first studies of H$_2$ gas at high redshifts have been done,
thanks to the gravitational telescopes. With the future mm instruments,
it will be possible to study the history of star formation, directly
with measuring the amount of gas available, and deriving the star
formation efficiency.

\end{article}
\end{document}